\documentclass[journal=jctcce,manuscript=communication]{achemso}
\setkeys{acs}{maxauthors=0,articletitle=true}

\usepackage{achemso}
\usepackage{graphics}
\usepackage{amssymb,amsfonts}
\usepackage{graphicx}
\usepackage[table,dvipsnames]{xcolor}
\usepackage{multirow}
\usepackage{caption}
\usepackage{subcaption}
\usepackage{booktabs}
\usepackage{colortbl}
\usepackage{amsmath}
\usepackage{amsopn}
\usepackage{bm}
\usepackage{siunitx}
\usepackage{bm}
\usepackage{color}
\usepackage{array}
\usepackage{lscape}
\usepackage{mciteplus}
\usepackage[version=3]{mhchem}
\usepackage{ulem}
\usepackage{listings}
\usepackage{enumerate}
\usepackage{lmodern}

\SectionNumbersOn
\DeclareMathOperator{\tr}{Tr}

\author{Janus J. Eriksen}
\email{janus@kemi.dtu.dk}
\affiliation{DTU Chemistry, Technical University of Denmark\\Kemitorvet Bldg. 206, DK--2800 Kgs. Lyngby, Denmark}

\title[TITLE]{Decomposed Mean-Field Simulations of Local Properties in Condensed Phases}

\begin{document}

%
%%%%%%%%%
%  ABSTRACT  %
%%%%%%%%%
%
\begin{abstract}

The present work demonstrates a robust protocol for probing localized electronic structure in condensed-phase systems, operating in terms of a recently proposed theory for decomposing the results of Kohn-Sham density functional theory in a basis of spatially localized molecular orbitals [Eriksen, {\textit{J. Chem. Phys.}} {\textbf{153}}, 214109 (2020)]. In an initial application to liquid, ambient water and the assessment of the solvation energy and the embedded dipole moment of H$_2$O in solution, we find that both properties are amplified on average---in accordance with expectation---and that correlations are indeed observed to exist between them. However, the simulated solvent-induced shift to the dipole moment of water is found to be significantly dampened with respect to typical literature values. The local nature of our methodology has further allowed us to evaluate the convergence of bulk properties with respect to the extent of the underlying one-electron basis set, ranging from single-$\zeta$ to full (augmented) quadruple-$\zeta$ quality. Albeit a pilot example, our work paves the way towards future studies of local effects and defects in more complex phases, e.g., liquid mixtures and even solid-state crystals. 

\end{abstract}

\newpage

%
%%%%%%%%%%%%
%    TOC GRAPHIC
%%%%%%%%%%%%

%
\section*{TOC Graphic}
\begin{figure}[ht!]
\begin{center}
\includegraphics[width=0.9\textwidth]{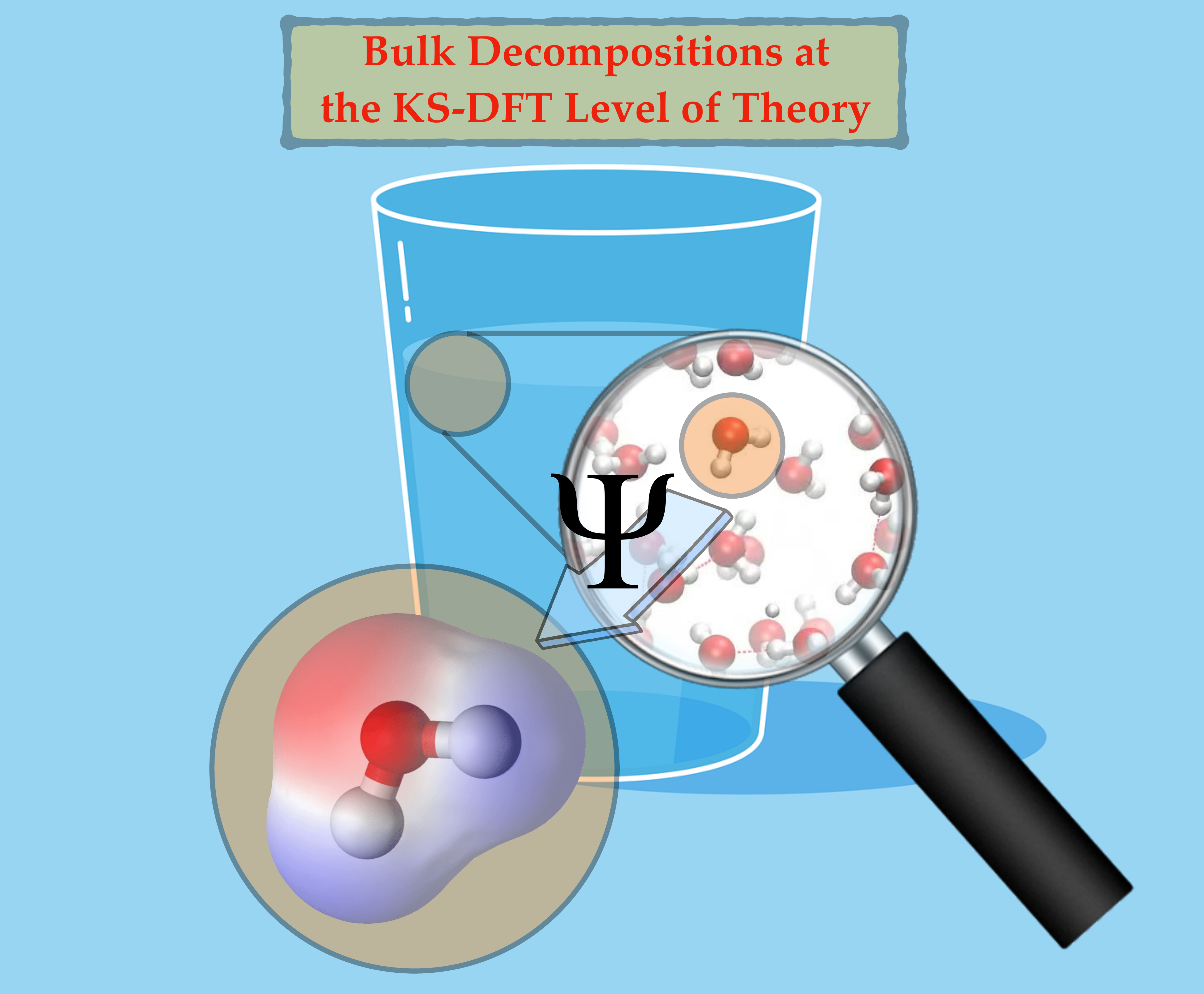}
\caption*{}
\label{toc_fig}
\end{center}
\end{figure}

\newpage

%
%%%%%%%%%%%
%  INTRODUCTION
%%%%%%%%%%%
%

Although the concept of local properties in extended bulk systems is easily contested as being ambiguous, its accurate and stringent determination by means of computational methodologies need not necessarily be. To that end, the properties of a single water molecule will be inherently different from those in liquid water, insofar as the electronic structure of an individual monomer in a droplet undoubtedly differs from that of its isolated counterpart due to charge transfer and delocalization effects. Divergences from isolated (vacuum) results will fundamentally be driven by perturbations to the local electronic structure of the embedded monomer, in addition to structural distortions mediated by the environment, and any consistent attempt at distinguishing individual property contributions to that of the bulk will inevitably prove difficult as a result. Although this will hold true for both the experimental and computational sides of things, the latter will benefit from being able to effectively interlink alterations to the electronic structure in the solvation process that drives the transition from a water molecule in isolation to what constitutes a minimal droplet model~\cite{ceotto_water_droplet_chem_sci_2021}.\\

If bulk properties are sought after as resolved into individual monomeric contributions, however, the employed methodology must partition the total quantities into a set of inherent terms that are subject to a number of requirements. First, local contributions must rely on objects that themselves remain spatially local, while simultaneously allowing for the effect of any nearby environment to fold into them. Second, the magnitude of a local (monomeric) property needs to stay satisfactorily invariant upon an enlargement of the underlying one-electron basis set---in addition to augmentations by diffuse functions---to make the involved simulations computationally tractable. Finally, the employed partitioning scheme should ideally stay sensitive enough to vary upon a refinement in the treatment of the electronic structure, if indeed correlation effects happen to be significant in the system at hand (which need not necessarily be the case). While the opposite would be desirable as well, the ability to encode an increased degree of physics into the computations and, in turn, benefit from this by having the added scrutiny reflected in the final results is, at the very least, theoretically appealing. We will here demonstrate this very point by presenting results for the solvation shifts to both monomer ground-state energies and dipole moments, as obtained at both the uncorrelated Hartree-Fock (HF) level and by using exchange-correlation ($xc$) functionals belonging to various rungs of the Kohn-Sham density functional theory (KS-DFT) hierarchy~\cite{perdew_jacobs_ladder_aip_conf_proc_2001}.\\

In a recent paper~\cite{eriksen_decodense_jcp_2020}, we set out to demonstrate how to decompose mean-field (MF) molecular energies and dipole moments by means of a spatially localized orbital basis. Importantly, the decomposition of these properties was mediated by dividing the full 1-electron reduced density matrix (1-RDM) into a set of contributions associated with either the specific bonds or atoms of a given system, without recourse to an explicit partitioning of the electron density. Specifically, Ref. \citenum{eriksen_decodense_jcp_2020} outlined how to decompose first-order electronic properties in a basis of atom-specific 1-RDMs, $\{\bm{\delta}\}$, which themselves are defined according to the definition:
\begin{align}
\bm{\delta}_K &= \sum_{\sigma}\bm{\delta}_{K,\sigma} = \sum_{\sigma}\sum^{\mathcal{N}_{\sigma}}_{i}\bm{d}_{i,\sigma}\bm{p}^{K}_{i,\sigma} \ . \label{atom_rdm1_eq}
\end{align}
In turn, the objects in Eq. \ref{atom_rdm1_eq} are defined via a set of 1-RDMs, $\bm{d}_{i,\sigma} = \bm{C}_{i,\sigma}\bm{C}^T_{i,\sigma}$, unique to the individual occupied spin-$\sigma$ molecular orbitals (MOs) of the system, $\bm{C}_{i,\sigma}$, and a set of population weights of all $\mathcal{N}_{\sigma}$ MOs of $\alpha$-/$\beta$-spin on a given atom $K$, $\{\bm{p}^{K}\}$. The earlier investigations in Ref. \citenum{eriksen_decodense_jcp_2020} clearly emphazised how the population weights used to assign $\{\bm{d}\}$ should ideally not be drawn from regular Mulliken population analyses~\cite{mulliken_population_jcp_1955}, but rather be recast into an alternative basis of reduced dimension, such as, the intrinsic AOs (IAOs) courtesy of Knizia~\cite{knizia_iao_ibo_jctc_2013}.\\

Using the atom-specific 1-RDMs in Eq. \ref{atom_rdm1_eq}, an MF energy may be reformulated into the following, partitioned form amongst the $\mathcal{M}_{\text{atom}}$ atoms of the system at hand
\begin{align}
E_{\text{MF}} = \sum^{\mathcal{M}_{\text{atom}}}_{K}E_{\text{elec},K}(\bm{D},\bm{\delta}_K) + (E_{xc,K}(\bm{\rho},\bm{\varrho}_K)) + E_{\text{nuc},K} \ , \label{atom_decomp_eq}
\end{align}
in terms of nuclear and electronic contributions that read as
\begin{subequations}
\label{atom_contr_eqs}
\begin{align}
E_{\text{nuc},K} &= Z_K\sum^{\mathcal{M}_{\text{atom}}}_{K>L}\frac{Z_L}{|\bm{r}_K - \bm{r}_L|} \label{nuc_contr_atom_eq} \\
E_{\text{elec},K} &= \tr[\bm{T}_{\text{kin}}\bm{\delta}_K] + \tfrac{1}{2}(\tr[\bm{V}_{K}\bm{D}] + \tr[\bm{V}_{\text{nuc}}\bm{\delta}_{K}]) + \tfrac{1}{2}\sum_{\sigma}\tr[\bm{G}_{\sigma}(\bm{D})\bm{\delta}_{K,\sigma}] \label{elec_contr_atom_eq} \\
E_{xc,K} &= \tr[\epsilon_{xc}(\bm{\rho})\bm{\varrho}_K] \ . \label{xc_contr_atom_eq}
\end{align}
\end{subequations}
In Eq. \ref{nuc_contr_atom_eq}, $Z_K$ and $\bm{r}_K$ denote the nuclear charge and position of atom $K$, while the kinetic energy and nuclear attraction operators in Eq. \ref{elec_contr_atom_eq} are denoted by $\bm{T}_{\text{kin}}$ and $\bm{V}_{\text{nuc}}$, respectively, alongside the attractive potential associated with atom $K$, $\bm{V}_{K}$, and an effective Fock potential, $\bm{G}_{\sigma}$. In Eq. \ref{elec_contr_atom_eq}, $\bm{D}$ denotes the full, spin-summed 1-RDM, while the $xc$ energy in Eq. \ref{xc_contr_atom_eq} is expressed in terms of the computed energy density, $\epsilon_{xc}$, as derived from the total electronic density, $\bm{\rho}$, and possibly its derivatives, which are quantities that may be trivially defined in an atom-specific manner, $\{\bm{\varrho}_K\}$, by proceeding through $\{\bm{\delta}_K\}$. Likewise, a molecular dipole moment---irrespective of the employed level of MF theory---may be expressed as
\begin{align}
\bm{\mu}_{\text{MF}} = \sum^{\mathcal{M}_{\text{atom}}}_{K}\bm{\mu}_{\text{elec},K}(\bm{\delta}_K) + \bm{\mu}_{\text{nuc},K} \ , \label{atom_decomp_dipmom_eq}
\end{align}
with nuclear and electronic contributions defined as
\begin{subequations}
\label{atom_contr_dipmom_eqs}
\begin{align}
\bm{\mu}^{\alpha}_{\text{nuc},K} &= Z_{K}\bm{r}^{\alpha}_{K} \label{nuc_contr_atom_dipmom_eq} \\
\bm{\mu}^{\alpha}_{\text{elec},K} &= -\tr[\bar{\bm{\mu}}^{\alpha}\bm{\delta}_{K}] \ , \label{elec_contr_atom_dipmom_eq}
\end{align}
\end{subequations}
in terms of AO dipole integrals, $\bar{\bm{\mu}}^{\alpha}$, for each of the Cartesian components, $\alpha=x,y,z$.\\

Provided with the theoretical setting above, a number of degrees of freedom in simulating a local bulk property still exist. The electronic contributions in Eqs. \ref{atom_contr_eqs} and \ref{atom_contr_dipmom_eqs} are fundamentally defined in terms of a spatially localized MO basis, so which choice should one ideally make for constructing this? In the context of the present study, we will use so-called intrinsic bond orbitals (IBOs)~\cite{knizia_iao_ibo_jctc_2013}, as defined by a standard Pipek-Mezey (PM) optimization scheme~\cite{pipek_mezey_jcp_1989}, but using atomic charges derived from IAO- rather than Mulliken-based populations~\cite{lehtola_jonsson_pm_jctc_2014}. In Figs. S1 and S2 of the Supporting Information (SI), results obtained using IBOs are compared to corresponding results obtained using either regular PM or Foster-Boys~\cite{foster_boys_rev_mod_phys_1960} (FB) localized MOs, in addition to results based on the equally lossless, but theoretically different energy density analysis (EDA) partitioning by Nakai~\cite{nakai_eda_partitioning_cpl_2002,nakai_eda_partitioning_ijqc_2009}, in which the total 1-RDM gets partitioned solely on the basis of which atoms the individual AOs are localized on (that is, irrespective of any further population measures). In our earlier study~\cite{eriksen_decodense_jcp_2020}, IBOs were found to be generally far superior to both FB and PM localized MOs, as these tended to yield unsystematic results for a selection of different systems. However, all three kinds of localized MOs are found to give comparable results in the present context, cf. Figs. S1 and S2 of the SI, which present results using either of the pc-1 and aug-pc-1 basis sets~\cite{jensen_pc_basis_sets_jcp_2001}, respectively. In fact, the consistency of the results obtained using any of the three choices of localized MOs increases in augmenting the pc-1 basis set by diffuse functions, whereas the opposite is pronouncedly true in the case of the EDA results, which are devoid of any ground for interpretation in the more realistic of the two basis sets. This discrepancy is due to the fact that the contributions yielded by EDA are intimately tied to the employed AOs rather than actual atoms, which, in turn, renders the decomposition somewhat predefined and insensitive to changes in the electronic structure. Add to that the observation that the EDA partitioning possesses no strict basis set limit, and we are ultimately left with IBOs (alongside IAOs for computing the weights in Eq. \ref{atom_rdm1_eq}) as a viable and rigorous option, and we will henceforth make exclusive use of these throughout the remainder of the present work.\\

In simulating local properties within a liquid bulk phase, one will need to sample dynamical effects both configurationally and radially by saturating results both with respect to fluctuations that come about as a result of temperature and the average local environment surrounding a central monomer unit. Simulating the actual bulk may be done by means of several treatments, e.g., force fields~\cite{paesani_water_ff_chem_rev_2016} or {\textit{ab initio}} molecular dynamics~\cite{tuckerman_aimd_perspective_pnas_2005} (AIMD), of which the latter constitutes the arguably most rigorous, albeit most costly option~\bibnote{In terms of accuracy, however, the data-driven many-body models from the Paesani group (described in Refs. \citenum{paesani_mb_pol_1_jctc_2013,paesani_mb_pol_2_jctc_2014,paesani_mb_pol_3_jctc_2014} and collectively reviewed in Ref. \citenum{paesani_water_ff_chem_rev_2016}) constitute the current state-of-the-art for water simulations, providing highly accurate descriptions of a wealth of properties in both the gas and condensed phase.}. The radial sampling of a local property is then carried out in a brute-force manner, by which a sphere surrounding a central monomer is incrementally enlarged until it encompasses a sufficient number of neighbours to warrant the intrinsic physics appropriately accounted for. Assuming that AIMD simulations have been properly equilibrated, the use of periodic boundary conditions (PBCs) renders the choice of central monomer to use further on relatively irrelevant. For consistency, however, it is usually wise to not only include different well-separated configurational snapshots in a sampling set, so as to avoid any autocorrelation between these, but also to extend the set of inputs by bulks centred around different monomers, which may be chosen upon at random. For samplings drawn from force-field simulations, it is reasonable just to focus on the monomer in closest proximity to the center of charge of the simulation body being employed. \\

\begin{figure}[ht!]
\vspace{-.6cm}
\begin{center}
\includegraphics[width=0.9\textwidth]{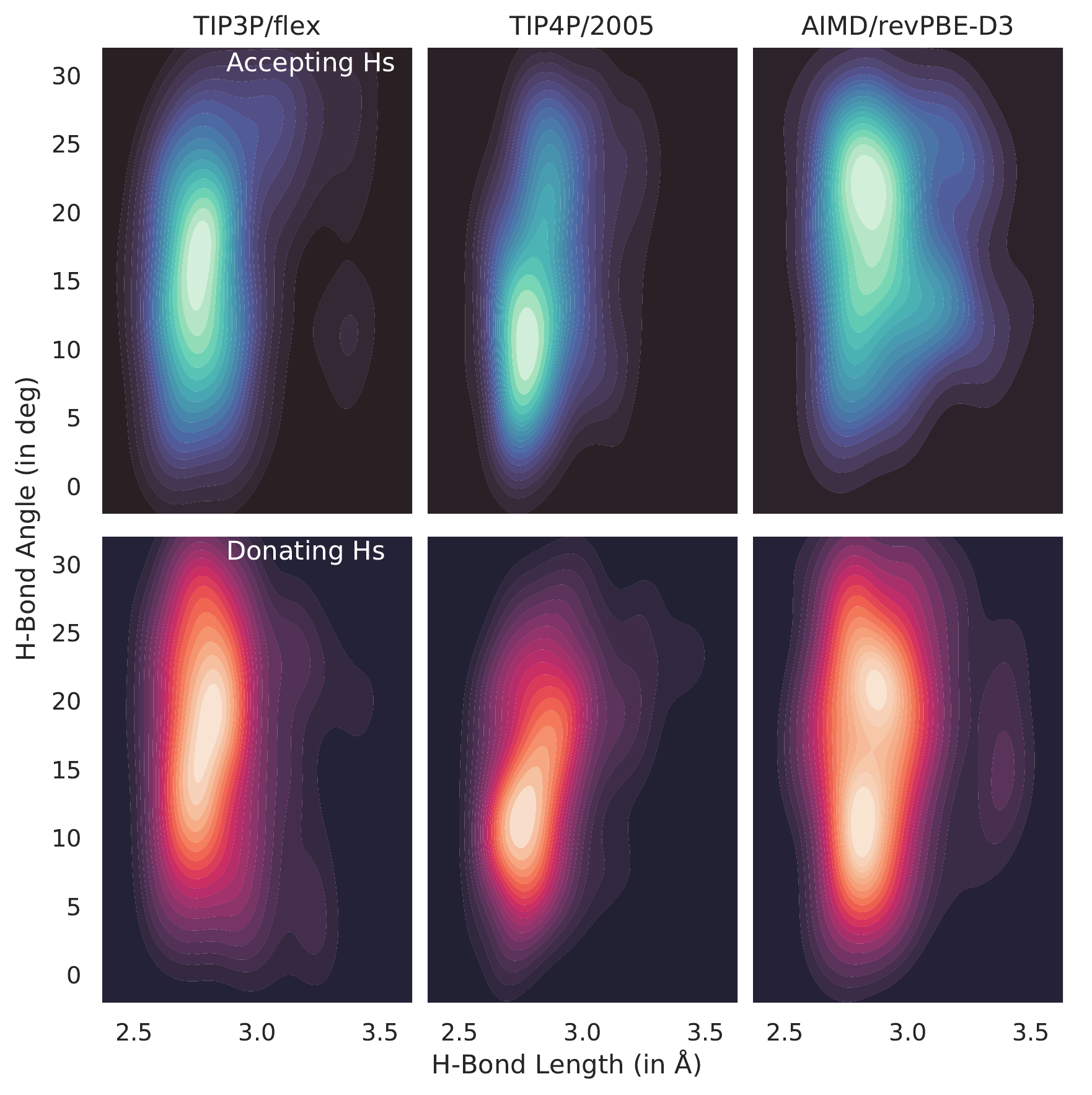}
\caption{Distributions of hydrogen bonds whereby a central monomer unit either accepts (upper panel) or donates (lower panel) a proton in the three different samplings. As per convention~\cite{luzar_chandler_water_h_bonds_prl_1996}, a hydrogen bond is defined in terms of a max. O--O distance of $3.5$ {\AA} and a max. angle between the donating O, donating H, and accepting O of $30^{\circ}$.}
\label{h_bonds_distrib_fig}
\end{center}
\vspace{-.8cm}
\end{figure}

In the present study, we will simulate the ground-state energy and molecular dipole moment of an embedded water monomer, as examples of prototypical local properties in a condensed phase, and we will base our study on 100 random snapshots from each of three fundamentally different samplings of liquid water: a simplified, yet flexible three-point TIP3P model~\cite{jorgensen_tip3p_jcp_1983} of a spherical droplet, a more advanced four-point TIP4P/2005 model~\cite{abascal_vega_tip4p_2005_jcp_2005} of a cubic box (with structural rigidity of the individual monomers enforced), and finally an AIMD simulation at the KS-DFT level of theory~\bibnote{In all three bulk samplings, the gauge origin of the AO dipole integrals have been fixed to coincide with the position of the central oxygen atom under investigation, which has further been translationally moved to the position, $\bm{R}_{\text{O}} = (0, 0, 0)$.}. All bulks are reasonably assumed equilibrated, and the reader is referred to the original Refs. \citenum{lilienfeld_fchl_jcp_2018,han_isborn_shi_water_dipole_jctc_2021,zhu_voorhis_water_dipole_jpcl_2021}, respectively, for further details on how the simulations were conducted. The hydrogen bonding a central monomer unit participates in is compared for the three different samplings in Fig. \ref{h_bonds_distrib_fig}, while Figs. S3 through S5 of the SI seek to compare these in terms of their bulk composition, fluctuations to their structural configurations, as well as their local orientational tetrahedral arrangements, respectively. As is evident from our comparisons of the bulks, quantifiable differences are indeed observed to exist between them, and it will hence prove instructive to study the extent to which these subtle, dynamical effects influence simulations of local properties within the liquid phase.\\

In terms of the electronic structure treatment, we will assess the quality of standard, restricted HF as well as a total of 4 different $xc$ functionals, namely, B3LYP~\cite{becke_b3lyp_functional_jcp_1993,frisch_b3lyp_functional_jpc_1994}, CAM-B3LYP~\cite{yanai_tew_handy_camb3lyp_functional_cpl_2004}, M06-2X~\cite{zhao_truhlar_m06_functional_tca_2008}, and $\omega$B97M-V~\cite{mardirossian_head_gordon_wb97m_v_functional_jcp_2016}. This selection of density functional approximations (DFAs) is motivated, in part, by the favoured candidates of the rigorous evaluation of KS-DFT for the simulation of dipole moments in Ref. \citenum{hait_head_gordon_dipole_mom_jctc_2018}. Of these, B3LYP is arguably the most popular, not only among its own class of (hybrid) generalized gradient approximations (GGAs), but across all of the available $xc$ functionals to date. The Coulomb-attenuated version of the functional, CAM-B3LYP, differs from the native B3LYP by implementing a mixing of short- and long-range exchange via the standard error function, while $\omega$B97M-V and M06-2X are meta-GGAs of the range-separated and global hybrid kind, respectively.\\

All KS-DFT calculations have been run with the {\texttt{PySCF}} program package~\cite{pyscf_wires_2018,pyscf_jcp_2020,libxc_software_x_2018}, while all subsequent decompositions have been performed using the {\texttt{decodense}} code~\cite{decodense}. The {\texttt{PySCF}} default number of (radial, angular Lebedev) grid points was used for the H and O atoms, i.e., $(50,302)$ and $(75,302)$, respectively, in combination with the standard pruning scheme, except for the double-grid integration involved in the evaluation of the nonlocal VV10 correlation for describing dispersion (van der Waals) interactions~\cite{vydrov_voorhis_vv10_functional_jcp_2010}, for which reduced SG-1 grids were used~\cite{gill_johnson_pople_sg1_cpl_1993}. The effect of employing denser quadrature grids was assessed for the $\omega$B97M-V and M06-2X functionals, as these involve complex expressions for the exchange inhomogeneity factor~\cite{dasgupta_herbert_sg2_sg3_jcc_2017}. For both DFAs, negligible errors were found, i.e., much smaller than those due to the use of a truncated radial extent (see below), and on par with errors due to the density fitting approximation~\cite{dunlap_densfit_pccp_2000}, which has similarly been invoked throughout due to its cost reductions.\\

Before embarking on simulating properties in each of the three bulks discussed above, we will first discuss the design of a suitable computational protocol. In Figs. S6 and S7 of the SI, results of a coarse-grained radial sampling extending outwards from a central monomer under consideration---at the B3LYP/(aug-)pc-1 level of theory---are observed to saturate at a radius well before the largest extension of $r=5.5$ {\AA}, a distance corresponding safely to the inclusion of the second solvation shell and a sphere encompassing in excess of 30 neighbouring monomers on average in all three samplings. No apparent differences in the convergence profiles are observed by adding diffuse functions. However, bulks of this extent are bound to render KS-DFT intractable in the enlarged and augmented basis sets that are compulsory to the determination of dipole moments. For this reason, we will here advocate in favour of the use of background point charges, which are observed from Figs. S6 and S7 to accelerate radial convergence---particularly in the case of dipole moments---and thus allow for reduced bulks of $r=4.0$ {\AA}, compromising the final accuracy of our simulations by only a fraction of a kcal/mol or a Debye in the case of energies and dipole moments, respectively~\bibnote{The TIP3P/flex and AIMD/revPBE-D3 samplings both use TIP3P charge distributions, namely, $q(\text{O}) = -0.834$ and $q(\text{H}) = 0.417$, while the TIP4P/2005 sampling makes use of its own three-point model, i.e., $q(\text{M}) = -1.1128$ and $q(\text{H}) = 0.5564$, with the negative charge placed on a dummy atom (M) at a distance of $0.1546$ {\AA} away from the oxygen along the $\angle$(H--O--H) bisector.}.\\
\begin{figure}[ht!]
\vspace{-.6cm}
\begin{center}
\includegraphics[width=0.9\textwidth]{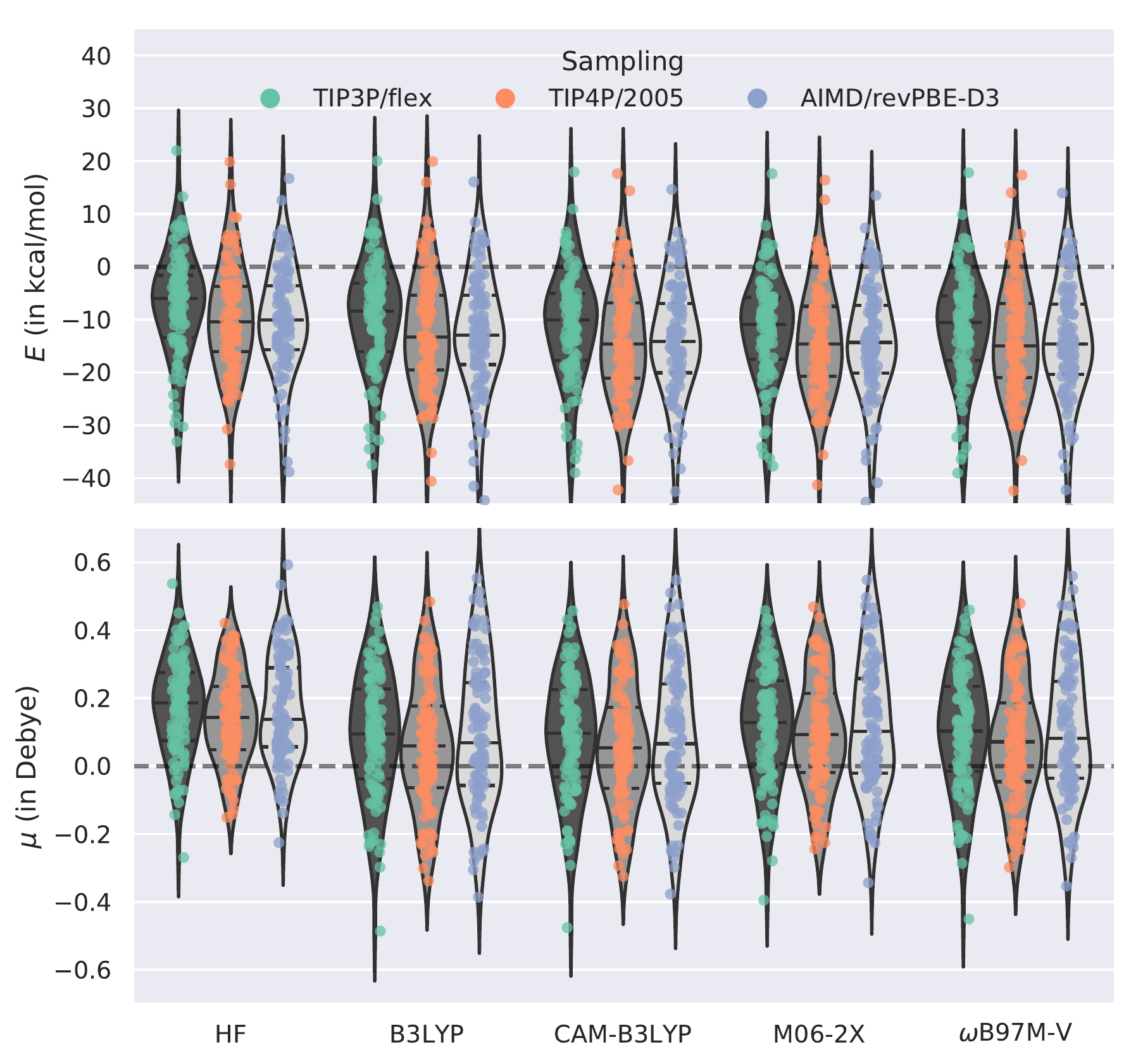}
\caption{Distributions of solvation energies (upper panel) and dipole moments (lower panel) for each of the three samplings and each of the four $xc$ functionals (using the aug-pc-1 basis set). In each violin, the quartiles of the underlying distribution have been displayed.}
\label{augpc1_fig}
\end{center}
\vspace{-.8cm}
\end{figure}

\begin{figure}[ht!]
\begin{center}
\includegraphics[width=0.9\textwidth]{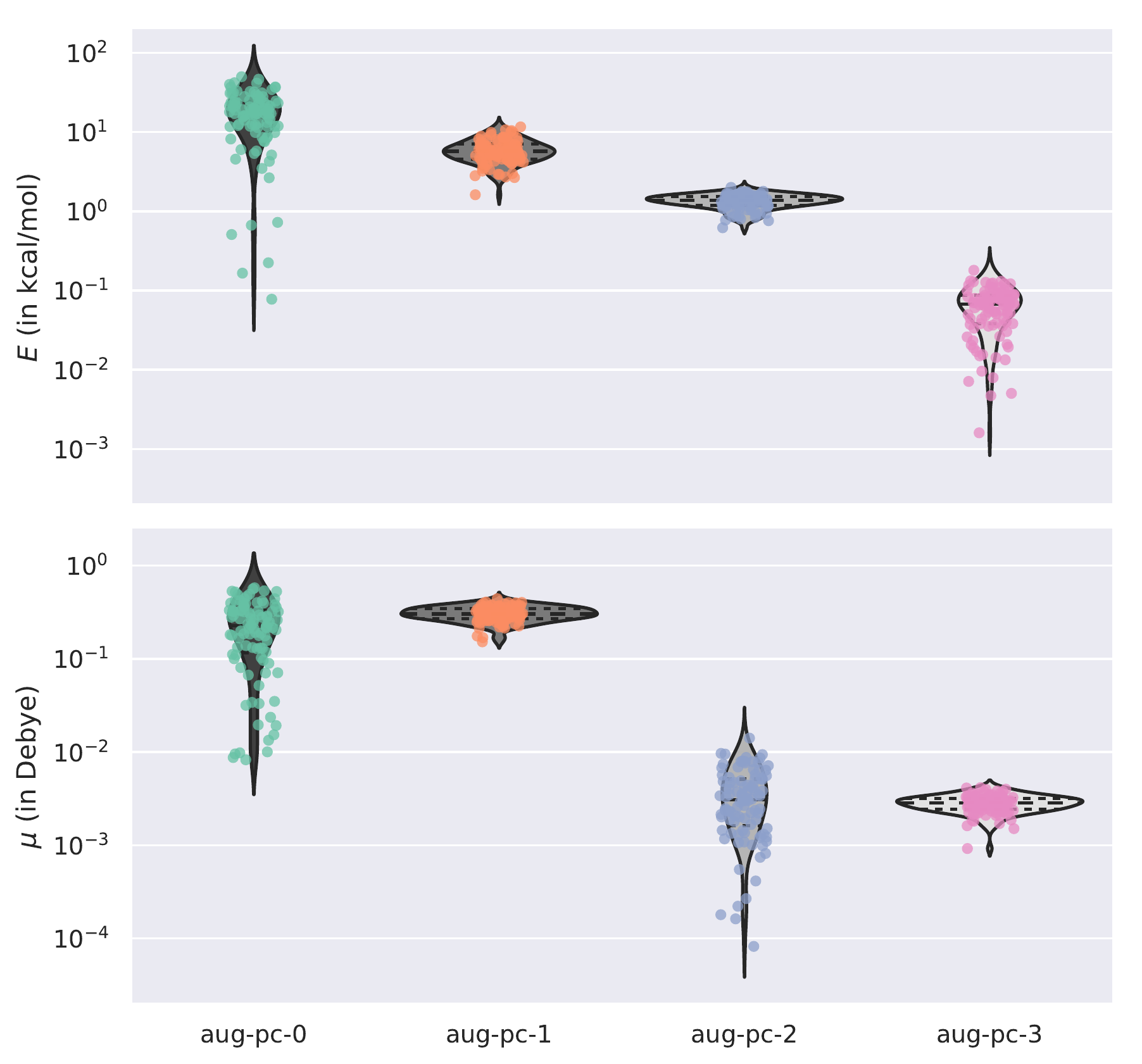}
\caption{Convergence of (absolute) solvation energies and dipole moments of H$_2$O in the AIMD/revPBE-D3 sampling, as calculated at the B3LYP/aug-pc-$x$ level of theory ($x=0$--$3$) and measured on a log-scale against results in the largest aug-pc-3 basis set.}
\label{aimd_b3lyp_basis_log_fig}
\end{center}
\vspace{-.6cm}
\end{figure}

Having established an efficient and functional protocol, the main results of the present study now follow. In Fig. \ref{augpc1_fig}, we report energies and molecular dipole moments, both measured as solvation shifts to corresponding vacuum results. In all three bulk samplings, HF and all of the DFAs under consideration have been evaluated in the aug-pc-1 basis set of double-$\zeta$ quality. From the results for both properties in Fig. \ref{augpc1_fig}, we observe little variation in-between the different bulks. At the same time, we note how the HF results slightly differ from those obtained using KS-DFT. Within any given bulk sampling, the individual distributions are difficult to recognize from one another, bar a minor potential shift, all giving mean solvation energies and dipole moments of approximately $-10$ kcal/mol and $0.1$ D, respectively, with a pronounced positive-valued tail for the latter property. Given these similarities, we may restrict our attention to only a single combination of DFA and bulk sampling in moving towards extended basis sets, e.g., the B3LYP $xc$ functional and the AIMD variant. Results in basis sets ranging from augmented single- to quadruple-$\zeta$ quality are presented in Fig. \ref{aimd_b3lyp_basis_log_fig}, which---besides verifying all conclusions based on the aug-pc-1 results in Fig. \ref{augpc1_fig}---reiterate the stable and rapid convergence of our decompositions with respect to basis set size first investigated in Ref. \citenum{eriksen_decodense_jcp_2020} (cf. also Fig. S8 of the SI, which presents Fig. \ref{aimd_b3lyp_basis_log_fig} on a linear scale).\\

\begin{figure}[ht!]
\begin{center}
\includegraphics[width=0.9\textwidth]{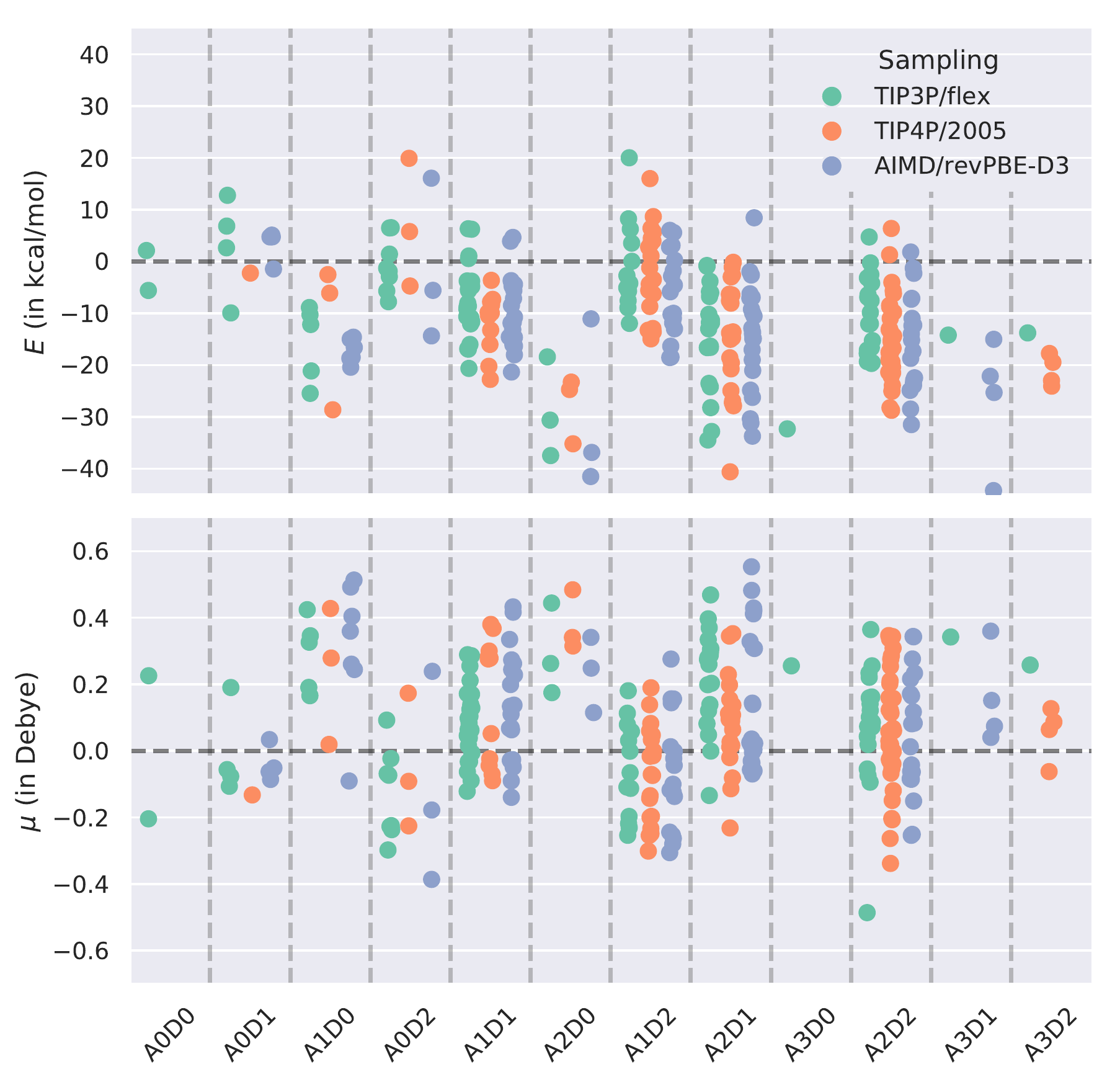}
\caption{Solvation energies and dipole moments of H$_2$O in all three samplings, as calculated at the B3LYP/aug-pc-1 level of theory. Results have been grouped on the basis of the number of protons accepted (A) and donated (D) by a central monomer in hydrogen bonds (Fig. \ref{h_bonds_distrib_fig}).}
\label{b3lyp_augpc1_h_bonds_class_fig}
\end{center}
\vspace{-.6cm}
\end{figure}

To inspect the findings of Figs. \ref{augpc1_fig} and \ref{aimd_b3lyp_basis_log_fig} in greater detail, we next group individual results on the basis of their local networks of hydrogen bonds, namely, the number of protons which a central monomer accepts and donates in these, cf. Fig. \ref{b3lyp_augpc1_h_bonds_class_fig}. In accordance with expectation, the monomers that accept more protons than they donate are generally stabilized more than those for which the opposite is true. This fact is also exemplified by Fig. S9 of the SI, which presents results for decomposed dipole moments along a constrained scan of the hydrogen bond in a simple water dimer. Further to that, a convincing pattern is observed to emerge for all three samplings in which shifts to ground-state energies and dipole moments are inversely proportional, in the sense that those monomers that energetically favour embedding also have net amplifications of their polar charge distributions. In support of this observation, these two simulated properties are plotted against one another in Fig. S10 of the SI where their signs and---to a somewhat lesser degree---their magnitudes are observed to correlate. However, what neither of Figs. \ref{b3lyp_augpc1_h_bonds_class_fig}, S9, and S10 succeed in quantifying is the extent to which a monomer is actually perturbed in a given configuration; nor do they fully show what the effects of potential asymmetries present in the hydrogen bond network surrounding said monomer amount to, nor how these factors explicitly correlate with the sign and magnitude of changes to the observed properties~\cite{bako_hermansson_water_dipole_jml_2019}. For this reason, all further attempts at unravelling and clarifying such relationships in bulk phases will be postponed to future studies.\\

\begin{figure}[ht!]
\begin{center}
\includegraphics[width=0.9\textwidth]{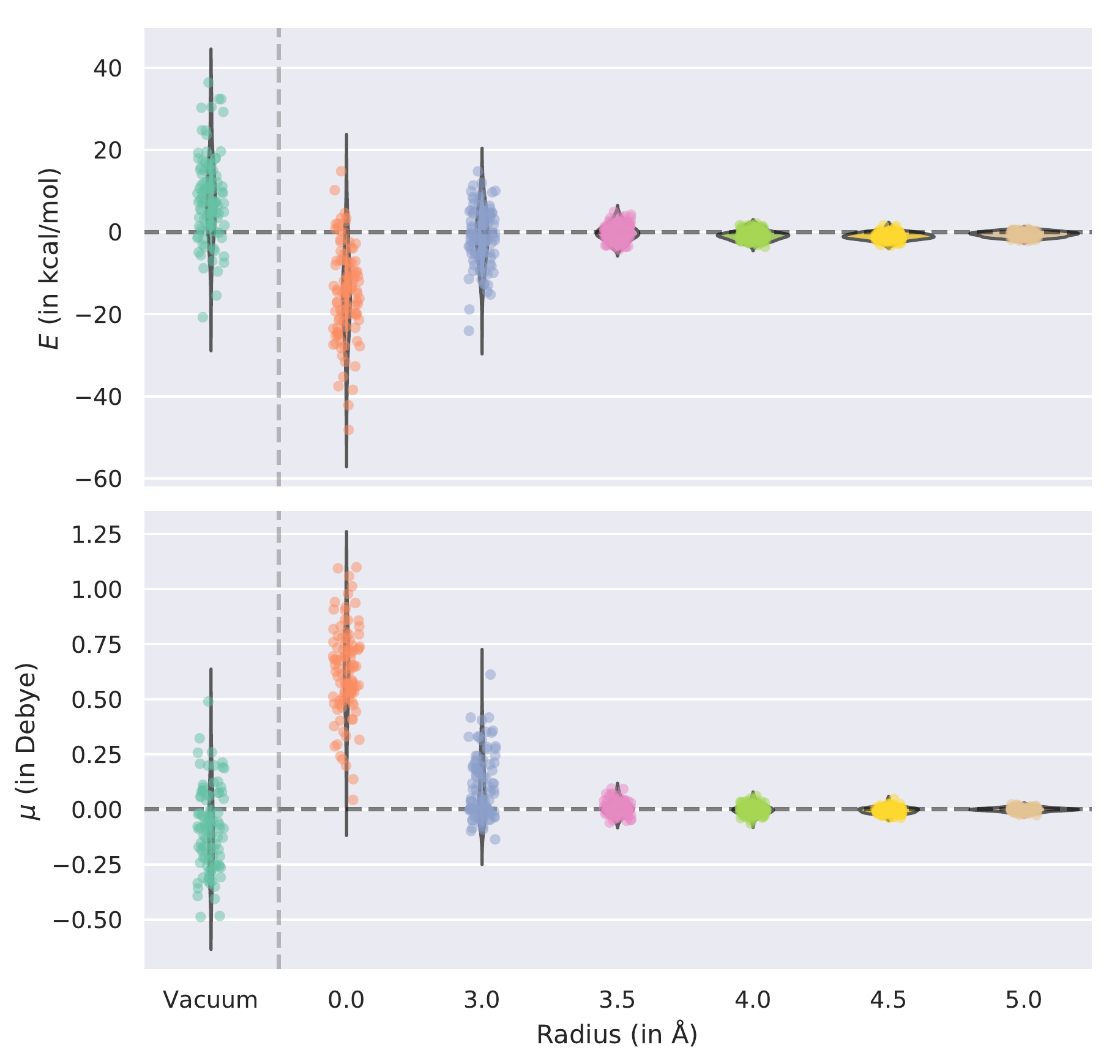}
\caption{Convergence of solvation energies and dipole moments of H$_2$O in the TIP3P/flex sampling against results at a radius of $r=5.5$ \AA, as calculated at the B3LYP/aug-pc-1 level of theory and simulated using background point charges, except where noted otherwise.}
\label{tip3p_flex_b3lyp_rad_conv_tot_fig}
\end{center}
\vspace{-.6cm}
\end{figure}

Finally, we note how we predict the overall (averaged) shift to the dipole moment of H$_2$O to be less than what has been reported by others, e.g., in the two recent studies by Han, Isborn, and Shi (Ref. \citenum{han_isborn_shi_water_dipole_jctc_2021}) and by Zhu and Van Voorhis (Ref. \citenum{zhu_voorhis_water_dipole_jpcl_2021}). In the former of these two, various schemes for assigning partial atomic charges (in response to local embedding environments) were evaluated, while in the latter, the authors employed the self-attractive Hartree (SAH) decomposition of the total electron density---as introduced in Ref. \citenum{zhu_silva_voorhis_sah_decomp_jctc_2018}---to simulate said dipole moments. Focusing on the latter study, the decomposition used in Ref. \citenum{zhu_voorhis_water_dipole_jpcl_2021} is characterized by a single, adjustable parameter that governs the degree of density localization, whereas the scheme used throughout the present work has a dependency on the local MOs and charge population proxy of choice. However, whereas practically no variance with respect to the employed MO basis is found herein, cf. Figs. S1 and S2 of the SI, the authors behind the SAH study found two distinct pictures, one localized and the other delocalized with less and more polarizable monomer dipoles, respectively, depending on the strength of their control parameter. Agreement between Ref. \citenum{zhu_voorhis_water_dipole_jpcl_2021} and the present study is met in the local limit, which yields the weakest solvation shifts, in our case by means of a protocol that remains spatially local in contrast to earlier attempts at employing localized MOs for this very purpose~\cite{bako_mayer_water_dipole_jpca_2016}. In Fig. \ref{tip3p_flex_b3lyp_rad_conv_tot_fig}, we show how too crude a truncation of the local environment in a simulation of local properties---particularly in combination with the use of background point charges~\cite{kongsted_water_dipole_cpl_2002,gordon_water_dipole_jpca_2008}---will lead to an overestimation of solvation effects, notably, in the case of dipole moments. As such, only upon saturating all possible charge (de)localization and polarization effects, even in a local picture frame, may reliable results be obtained~\cite{bako_mayer_water_dipole_jpcb_2016}, and these properties happen to differ only very moderately from results for isolated monomers in the averaged, rather than instantaneous limit. This also serves as a warning of how the common practice of extending on conclusions drawn from smaller, idealized water clusters to the bulk limit may be inherently problematic~\cite{gregory_saykally_water_dipole_science_1997}, as local properties are anything but insensitive to the heterogeneous local environment of water and its evolution over time. Unlike the situation in ice~\cite{coulson_eisenbergf_ice_dipole_proc_soc_london_1966,xantheas_jonsson_ice_dipole_jcp_1998}, networks of hydrogen bonds in ambient water are disordered, exhibiting a dynamical mixture of various identifiable configurations~\cite{liu_he_zhang_water_h_bonds_pccp_2017,head_gordon_water_aimd_chem_sci_2017,paesani_water_h_bonds_chem_sci_2019}, and the nature of molecular dipoles in solution will necessarily depend on this disorder in complex manners~\cite{torii_water_dipole_jpca_2013}.\\

In summary, we have demonstrated how an atom-centric decomposition theory for partitioning mean-field properties in a basis of spatially localized MOs may be leveraged as a robust protocol for probing localized electronic structure in condensed-phase systems~\cite{eriksen_decodense_jcp_2020}, with an initial application to solvation properties in liquid water. We find that results for both energies and dipole moments in the bulk phase are pronouncedly invariant, not only with respect to the sophistication of the underlying sampling procedure, but also to that of the employed DFA and the involved level of theory. Importantly, we have been able to move beyond the prevailing use of modest basis sets, restating the stable and rapid convergence of our decompositions through basis sets of augmented quadruple-$\zeta$ quality. These results have verified our in-depth analyses in the pragmatic, but in our case sufficient, aug-pc-1 basis set.\\

It is difficult to ascribe divergences of our simulated dipole moments from literature values, e.g., obtained by means of experimental X-ray diffraction techniques ($\mu = 2.9\pm0.6$ D)~\cite{badyal_soper_water_exp_dipole_jcp_2000}, as being caused by missing accounts of residual electron correlation, since all of our tested DFAs agree with one another, despite their functional differences. In addition, even experimental determinations of the monomer dipole in liquid water are not free of ambiguities, as these will all rely on some kind of fitting procedure to a model for the involved charge transfer processes, unlike in vacuum ($\mu = 1.855$ D) where Stark effect measurements can determine equilibrium dipole moments~\cite{clough_rothman_water_exp_dipole_jcp_1973}. Instead, both the assessment and interpretation of this and related properties in solution may need be reevaluated. Measured in close proximity of the isolated monomer alone, the dipole moment of water is clearly enhanced in the condensed phase. At a distance, however, its solvation shift is consistently observed to be significantly dampened due to an effective averaging of effects related to the relocation of electronic density. This is further motivated by the fact that our final solvation energies compare nicely with the enthalpy of vaporization of liquid water (ca. $10.5$ kcal/mol at \SI{25}{\celsius})~\cite{crc_85}. We will here argue that the present results---given the convincing saturation with respect to both electron correlation and basis set extension---provide a unique view on the dynamical interplay of individual monomers in liquid, ambient water, with potential applications of interest in data-driven models~\cite{unke_meuwly_ml_physnet_jctc_2019,ceriotti_ml_dipole_jcp_2020} and as a training pool in the design of new force fields~\cite{pande_water_ff_jpcb_2013,paesani_water_ff_jcp_2020}.

%
%%%%%%%%%%%%%%%
%  ACKNOWLEDGMENTS 
%%%%%%%%%%%%%%%
%
\section*{Acknowledgments}

This work was supported by a generous research grant (no. 37411) from VILLUM FONDEN (a part of THE VELUX FOUNDATIONS). Preliminary work was carried out while the author was still based at the School of Chemistry, University of Bristol, and The Independent Research Fund Denmark is gratefully acknowledged for financial support throughout this early phase, during which work was carried out using the computational facilities of the Advanced Computing Research Centre, University of Bristol. Finally, the author wishes to thank Liang Shi (University of California, Merced) and Tianyu Zhu (Caltech) for sharing the TIP4P/2005 and AIMD/revPBE-D3 samplings of Ref. \citenum{han_isborn_shi_water_dipole_jctc_2021} and \citenum{zhu_voorhis_water_dipole_jpcl_2021}, respectively.

%
%%%%%%%%%%%%%%%%%%
%  SUPPORTING INFORMATION
%%%%%%%%%%%%%%%%%%
%
\section*{Supporting Information}

The Supporting Information collects a number of additional results in support of the main results reported in the present Letter. The use of IBOs, PM, and FB localized MOs are compared in Figs. S1 and S2 (also in comparison to Nakai's EDA partitioning), in the pc-1 and aug-pc-1 basis sets, respectively. Fig. S3 depicts the radial distributions of neighbouring monomers within each of the three samplings, while Fig. S4 depicts the distributions of bond lengths and angles for the central monomers within the samplings. Fig. S5 quantifies the distributions of hydrogen bonds in the three sampling through results for the corresponding orientational tetrahedral orders to emphasize any deviations from regular tetrahedral configurations. The convergences of results obtained for the energy and dipole moment of an embedded water monomer with outwards distance are evaluated in Figs. S6 and S7, again for each of the pc-1 and aug-pc-1 basis sets, respectively, and Fig. S8 presents the basis set convergence results behind Fig. \ref{aimd_b3lyp_basis_log_fig} plotted on a linear scale, rather than as absolute numbers on a log-scale. Finally, results for decomposed molecular dipole moments along an optimized (constrained) potential energy scan of the water dimer are presented in Fig. S9, while the two properties in question are finally correlated against one another in Fig. S10.

\newpage

\providecommand{\latin}[1]{#1}
\makeatletter
\providecommand{\doi}
  {\begingroup\let\do\@makeother\dospecials
  \catcode`\{=1 \catcode`\}=2 \doi@aux}
\providecommand{\doi@aux}[1]{\endgroup\texttt{#1}}
\makeatother
\providecommand*\mcitethebibliography{\thebibliography}
\csname @ifundefined\endcsname{endmcitethebibliography}
  {\let\endmcitethebibliography\endthebibliography}{}


\begin{mcitethebibliography}{55}
\providecommand*\natexlab[1]{#1}
\providecommand*\mciteSetBstSublistMode[1]{}
\providecommand*\mciteSetBstMaxWidthForm[2]{}
\providecommand*\mciteBstWouldAddEndPuncttrue
  {\def\EndOfBibitem{\unskip.}}
\providecommand*\mciteBstWouldAddEndPunctfalse
  {\let\EndOfBibitem\relax}
\providecommand*\mciteSetBstMidEndSepPunct[3]{}
\providecommand*\mciteSetBstSublistLabelBeginEnd[3]{}
\providecommand*\EndOfBibitem{}
\mciteSetBstSublistMode{f}
\mciteSetBstMaxWidthForm{subitem}{(\alph{mcitesubitemcount})}
\mciteSetBstSublistLabelBeginEnd
  {\mcitemaxwidthsubitemform\space}
  {\relax}
  {\relax}

\bibitem[Rognoni \latin{et~al.}(2021)Rognoni, Conte, and
  Ceotto]{ceotto_water_droplet_chem_sci_2021}
Rognoni,~A.; Conte,~R.; Ceotto,~M. {How Many Water Molecules are Needed to
  Solvate One?} \emph{Chem. Sci.} \textbf{2021}, \emph{12}, 2060\relax
\mciteBstWouldAddEndPuncttrue
\mciteSetBstMidEndSepPunct{\mcitedefaultmidpunct}
{\mcitedefaultendpunct}{\mcitedefaultseppunct}\relax
\EndOfBibitem
\bibitem[Perdew and Schmidt(2001)Perdew, and
  Schmidt]{perdew_jacobs_ladder_aip_conf_proc_2001}
Perdew,~J.~P.; Schmidt,~K. {Jacob's Ladder of Density Functional Approximations
  for the Exchange-Correlation Energy}. \emph{AIP Conf. Proc.} \textbf{2001},
  \emph{577}, 1\relax
\mciteBstWouldAddEndPuncttrue
\mciteSetBstMidEndSepPunct{\mcitedefaultmidpunct}
{\mcitedefaultendpunct}{\mcitedefaultseppunct}\relax
\EndOfBibitem
\bibitem[Eriksen(2020)]{eriksen_decodense_jcp_2020}
Eriksen,~J.~J. {Mean-Field Density Matrix Decompositions}. \emph{{J}. {C}hem.
  {P}hys.} \textbf{2020}, \emph{153}, 214109\relax
\mciteBstWouldAddEndPuncttrue
\mciteSetBstMidEndSepPunct{\mcitedefaultmidpunct}
{\mcitedefaultendpunct}{\mcitedefaultseppunct}\relax
\EndOfBibitem
\bibitem[Mulliken(1955)]{mulliken_population_jcp_1955}
Mulliken,~R.~S. {Electronic Population Analysis on LCAO-MO Molecular Wave
  Functions. I}. \emph{J. Chem. Phys.} \textbf{1955}, \emph{23}, 1833\relax
\mciteBstWouldAddEndPuncttrue
\mciteSetBstMidEndSepPunct{\mcitedefaultmidpunct}
{\mcitedefaultendpunct}{\mcitedefaultseppunct}\relax
\EndOfBibitem
\bibitem[Knizia(2013)]{knizia_iao_ibo_jctc_2013}
Knizia,~G. {Intrinsic Atomic Orbitals: An Unbiased Bridge Between Quantum
  Theory and Chemical Concepts}. \emph{{J}. {C}hem. {T}heory {C}omput.}
  \textbf{2013}, \emph{9}, 4834\relax
\mciteBstWouldAddEndPuncttrue
\mciteSetBstMidEndSepPunct{\mcitedefaultmidpunct}
{\mcitedefaultendpunct}{\mcitedefaultseppunct}\relax
\EndOfBibitem
\bibitem[Pipek and Mezey(1989)Pipek, and Mezey]{pipek_mezey_jcp_1989}
Pipek,~J.; Mezey,~P.~G. {A Fast Intrinsic Localization Procedure Applicable for
  {\it{Ab Initio}} and Semiempirical Linear Combination of Atomic Orbital Wave
  Functions}. \emph{{J}. {C}hem. {P}hys.} \textbf{1989}, \emph{90}, 4916\relax
\mciteBstWouldAddEndPuncttrue
\mciteSetBstMidEndSepPunct{\mcitedefaultmidpunct}
{\mcitedefaultendpunct}{\mcitedefaultseppunct}\relax
\EndOfBibitem
\bibitem[Lehtola and J{\'o}nsson(2014)Lehtola, and
  J{\'o}nsson]{lehtola_jonsson_pm_jctc_2014}
Lehtola,~S.; J{\'o}nsson,~H. {Pipek-Mezey Orbital Localization Using Various
  Partial Charge Estimates}. \emph{{J}. {C}hem. {T}heory {C}omput.}
  \textbf{2014}, \emph{10}, 642\relax
\mciteBstWouldAddEndPuncttrue
\mciteSetBstMidEndSepPunct{\mcitedefaultmidpunct}
{\mcitedefaultendpunct}{\mcitedefaultseppunct}\relax
\EndOfBibitem
\bibitem[Foster and Boys(1960)Foster, and Boys]{foster_boys_rev_mod_phys_1960}
Foster,~J.~M.; Boys,~S.~F. {Canonical Configurational Interaction Procedure}.
  \emph{{R}ev. {M}od. {P}hys.} \textbf{1960}, \emph{32}, 300\relax
\mciteBstWouldAddEndPuncttrue
\mciteSetBstMidEndSepPunct{\mcitedefaultmidpunct}
{\mcitedefaultendpunct}{\mcitedefaultseppunct}\relax
\EndOfBibitem
\bibitem[Nakai(2002)]{nakai_eda_partitioning_cpl_2002}
Nakai,~H. {Energy Density Analysis with Kohn-Sham Orbitals}. \emph{Chem. Phys.
  Lett.} \textbf{2002}, \emph{363}, 73\relax
\mciteBstWouldAddEndPuncttrue
\mciteSetBstMidEndSepPunct{\mcitedefaultmidpunct}
{\mcitedefaultendpunct}{\mcitedefaultseppunct}\relax
\EndOfBibitem
\bibitem[Kikuchi \latin{et~al.}(2009)Kikuchi, Imamura, and
  Nakai]{nakai_eda_partitioning_ijqc_2009}
Kikuchi,~Y.; Imamura,~Y.; Nakai,~H. {One-Body Energy Decomposition Schemes
  Revisited: Assessment of Mulliken-, Grid-, and Conventional Energy Density
  Analyses}. \emph{Int. J. Quantum Chem.} \textbf{2009}, \emph{109}, 2464\relax
\mciteBstWouldAddEndPuncttrue
\mciteSetBstMidEndSepPunct{\mcitedefaultmidpunct}
{\mcitedefaultendpunct}{\mcitedefaultseppunct}\relax
\EndOfBibitem
\bibitem[Jensen(2001)]{jensen_pc_basis_sets_jcp_2001}
Jensen,~F. {Polarization Consistent Basis Sets: Principles}. \emph{{J}. {C}hem.
  {P}hys.} \textbf{2001}, \emph{115}, 9113\relax
\mciteBstWouldAddEndPuncttrue
\mciteSetBstMidEndSepPunct{\mcitedefaultmidpunct}
{\mcitedefaultendpunct}{\mcitedefaultseppunct}\relax
\EndOfBibitem
\bibitem[Cisneros \latin{et~al.}(2016)Cisneros, Wikfeldt, Ojam{\"a}e, Lu, Xu,
  Torabifard, Bart{\'o}k, Cs{\'a}nyi, Molinero, and
  Paesani]{paesani_water_ff_chem_rev_2016}
Cisneros,~G.~A.; Wikfeldt,~K.~T.; Ojam{\"a}e,~L.; Lu,~J.; Xu,~Y.;
  Torabifard,~H.; Bart{\'o}k,~A.~P.; Cs{\'a}nyi,~G.; Molinero,~V.; Paesani,~F.
  {Modeling Molecular Interactions in Water: From Pairwise to Many-Body
  Potential Energy Functions}. \emph{Chem. Rev.} \textbf{2016}, \emph{116},
  7501\relax
\mciteBstWouldAddEndPuncttrue
\mciteSetBstMidEndSepPunct{\mcitedefaultmidpunct}
{\mcitedefaultendpunct}{\mcitedefaultseppunct}\relax
\EndOfBibitem
\bibitem[Iftimie \latin{et~al.}(2005)Iftimie, Minary, and
  Tuckerman]{tuckerman_aimd_perspective_pnas_2005}
Iftimie,~R.; Minary,~P.; Tuckerman,~M.~E. {{\textit{Ab Initio}} Molecular
  Dynamics: Concepts, Recent Developments, and Future Trends}. \emph{Proc.
  Natl. Acad. Sci.} \textbf{2005}, \emph{102}, 6654\relax
\mciteBstWouldAddEndPuncttrue
\mciteSetBstMidEndSepPunct{\mcitedefaultmidpunct}
{\mcitedefaultendpunct}{\mcitedefaultseppunct}\relax
\EndOfBibitem
\bibitem[Not()]{Note-1}
In terms of accuracy, however, the data-driven many-body models
from the Paesani group (described in Refs.
\citenum{paesani_mb_pol_1_jctc_2013,paesani_mb_pol_2_jctc_2014,paesani_mb_pol_3_jctc_2014}
and collectively reviewed in Ref. \citenum{paesani_water_ff_chem_rev_2016})
constitute the current state-of-the-art for water simulations, providing highly
accurate descriptions of a wealth of properties in both the gas and condensed phase.\relax
\mciteBstWouldAddEndPunctfalse
\mciteSetBstMidEndSepPunct{\mcitedefaultmidpunct}
{}{\mcitedefaultseppunct}\relax
\EndOfBibitem
\bibitem[Babin \latin{et~al.}(2013)Babin, Leforestier, and Paesani]{paesani_mb_pol_1_jctc_2013}
Babin,~V.; Leforestier,~C.; Paesani,~F.
  {Development of a ``First Principles" Water Potential with Flexible Monomers:
  Dimer Potential Energy Surface, VRT Spectrum, and Second Virial Coefficient}.
  \emph{J. Chem. Theory Comput.} \textbf{2013}, \emph{9}, 5395\relax
\mciteBstWouldAddEndPuncttrue
\mciteSetBstMidEndSepPunct{\mcitedefaultmidpunct}
{\mcitedefaultendpunct}{\mcitedefaultseppunct}\relax
\EndOfBibitem
\bibitem[Babin \latin{et~al.}(2014)Babin, Medders, and Paesani]{paesani_mb_pol_2_jctc_2014}
Babin,~V.; Medders,~G.~R.; Paesani,~F.
  {Development of a ``First Principles" Water Potential with Flexible Monomers.
  II: Trimer Potential Energy Surface, Third Virial Coefficient, and Small Clusters}.
  \emph{J. Chem. Theory Comput.} \textbf{2014}, \emph{10}, 1599\relax
\mciteBstWouldAddEndPuncttrue
\mciteSetBstMidEndSepPunct{\mcitedefaultmidpunct}
{\mcitedefaultendpunct}{\mcitedefaultseppunct}\relax
\EndOfBibitem
\bibitem[Medders \latin{et~al.}(2014)Medders, Babin, and Paesani]{paesani_mb_pol_3_jctc_2014}
Medders,~G.~R.; Babin,~V.; Paesani,~F.
  {Development of a ``First Principles" Water Potential with Flexible Monomers.
  III: Liquid Phase Properties}.
  \emph{J. Chem. Theory Comput.} \textbf{2014}, \emph{10}, 2906\relax
\mciteBstWouldAddEndPuncttrue
\mciteSetBstMidEndSepPunct{\mcitedefaultmidpunct}
{\mcitedefaultendpunct}{\mcitedefaultseppunct}\relax
\EndOfBibitem
\bibitem[Jorgensen \latin{et~al.}(1983)Jorgensen, Chandrasekhar, Madura, Impey,
  and Klein]{jorgensen_tip3p_jcp_1983}
Jorgensen,~W.~L.; Chandrasekhar,~J.; Madura,~J.~D.; Impey,~R.~W.; Klein,~M.~L.
  {Comparison of Simple Potential Functions for Simulating Liquid Water}.
  \emph{J. Chem. Phys.} \textbf{1983}, \emph{79}, 926\relax
\mciteBstWouldAddEndPuncttrue
\mciteSetBstMidEndSepPunct{\mcitedefaultmidpunct}
{\mcitedefaultendpunct}{\mcitedefaultseppunct}\relax
\EndOfBibitem
\bibitem[Abascal and Vega(2005)Abascal, and
  Vega]{abascal_vega_tip4p_2005_jcp_2005}
Abascal,~J. L.~F.; Vega,~C. {A General Purpose Model for the Condensed Phases
  of Water: TIP4P/2005}. \emph{J. Chem. Phys.} \textbf{2005}, \emph{123},
  234505\relax
\mciteBstWouldAddEndPuncttrue
\mciteSetBstMidEndSepPunct{\mcitedefaultmidpunct}
{\mcitedefaultendpunct}{\mcitedefaultseppunct}\relax
\EndOfBibitem
\bibitem[Not()]{Note-2}
In all three bulk samplings, the gauge origin of the AO dipole integrals have
  been fixed to coincide with the position of the central oxygen atom under
  investigation, which has further been translationally moved to the position,
  $\bm{R}_{\text{O}} = (0, 0, 0)$.\relax
\mciteBstWouldAddEndPunctfalse
\mciteSetBstMidEndSepPunct{\mcitedefaultmidpunct}
{}{\mcitedefaultseppunct}\relax
\EndOfBibitem
\bibitem[Faber \latin{et~al.}(2018)Faber, Christensen, Huang, and von
  Lilienfeld]{lilienfeld_fchl_jcp_2018}
Faber,~F.~A.; Christensen,~A.~S.; Huang,~B.; von Lilienfeld,~O.~A. {Alchemical
  and Structural Distribution Based Representation for Universal Quantum
  Machine Learning}. \emph{J. Chem. Phys.} \textbf{2018}, \emph{148},
  241717\relax
\mciteBstWouldAddEndPuncttrue
\mciteSetBstMidEndSepPunct{\mcitedefaultmidpunct}
{\mcitedefaultendpunct}{\mcitedefaultseppunct}\relax
\EndOfBibitem
\bibitem[Han \latin{et~al.}(2021)Han, Isborn, and
  Shi]{han_isborn_shi_water_dipole_jctc_2021}
Han,~B.; Isborn,~C.~M.; Shi,~L. {Determining Partial Atomic Charges for Liquid
  Water: Assessing Electronic Structure and Charge Models}. \emph{J. Chem.
  Theory Comput.} \textbf{2021}, \emph{17}, 889\relax
\mciteBstWouldAddEndPuncttrue
\mciteSetBstMidEndSepPunct{\mcitedefaultmidpunct}
{\mcitedefaultendpunct}{\mcitedefaultseppunct}\relax
\EndOfBibitem
\bibitem[Zhu and Van~Voorhis(2021)Zhu, and
  Van~Voorhis]{zhu_voorhis_water_dipole_jpcl_2021}
Zhu,~T.; Van~Voorhis,~T. {Understanding the Dipole Moment of Liquid Water from
  a Self-Attractive Hartree Decomposition}. \emph{J. Phys. Chem. Lett.}
  \textbf{2021}, \emph{12}, 6\relax
\mciteBstWouldAddEndPuncttrue
\mciteSetBstMidEndSepPunct{\mcitedefaultmidpunct}
{\mcitedefaultendpunct}{\mcitedefaultseppunct}\relax
\EndOfBibitem
\bibitem[Luzar and Chandler(1996)Luzar, and
  Chandler]{luzar_chandler_water_h_bonds_prl_1996}
Luzar,~A.; Chandler,~D. {Effect of Environment on Hydrogen Bond Dynamics in
  Liquid Water}. \emph{Phys. Rev. Lett.} \textbf{1996}, \emph{76}, 928\relax
\mciteBstWouldAddEndPuncttrue
\mciteSetBstMidEndSepPunct{\mcitedefaultmidpunct}
{\mcitedefaultendpunct}{\mcitedefaultseppunct}\relax
\EndOfBibitem
\bibitem[Becke(1993)]{becke_b3lyp_functional_jcp_1993}
Becke,~A.~D. {Density-Functional Thermochemistry. III. The Role of Exact
  Exchange}. \emph{J. Chem. Phys.} \textbf{1993}, \emph{98}, 5648\relax
\mciteBstWouldAddEndPuncttrue
\mciteSetBstMidEndSepPunct{\mcitedefaultmidpunct}
{\mcitedefaultendpunct}{\mcitedefaultseppunct}\relax
\EndOfBibitem
\bibitem[Stephens \latin{et~al.}(1994)Stephens, Devlin, Chabalowski, and
  Frisch]{frisch_b3lyp_functional_jpc_1994}
Stephens,~P.~J.; Devlin,~F.~J.; Chabalowski,~C.~F.; Frisch,~M.~J. {{\it{Ab
  Initio}} Calculation of Vibrational Absorption and Circular Dichroism Spectra
  Using Density Functional Force Fields}. \emph{J. Phys. Chem.} \textbf{1994},
  \emph{98}, 11623\relax
\mciteBstWouldAddEndPuncttrue
\mciteSetBstMidEndSepPunct{\mcitedefaultmidpunct}
{\mcitedefaultendpunct}{\mcitedefaultseppunct}\relax
\EndOfBibitem
\bibitem[Yanai \latin{et~al.}(2004)Yanai, Tew, and
  Handy]{yanai_tew_handy_camb3lyp_functional_cpl_2004}
Yanai,~T.; Tew,~D.~P.; Handy,~N.~C. {A New Hybrid Exchange-Correlation
  Functional Using the Coulomb-Attenuating Method (CAM-B3LYP)}. \emph{Chem.
  Phys. Lett.} \textbf{2004}, \emph{393}, 51\relax
\mciteBstWouldAddEndPuncttrue
\mciteSetBstMidEndSepPunct{\mcitedefaultmidpunct}
{\mcitedefaultendpunct}{\mcitedefaultseppunct}\relax
\EndOfBibitem
\bibitem[Zhao and Truhlar(2008)Zhao, and
  Truhlar]{zhao_truhlar_m06_functional_tca_2008}
Zhao,~Y.; Truhlar,~D.~G. {The M06 Suite of Density Functionals for Main Group
  Thermochemistry, Thermochemical Kinetics, Noncovalent Interactions, Excited
  States, and Transition Elements: Two New Functionals and Systematic Testing
  of Four M06-Class Functionals and 12 Other Functionals}. \emph{Theor. Chem.
  Acc.} \textbf{2008}, \emph{120}, 215\relax
\mciteBstWouldAddEndPuncttrue
\mciteSetBstMidEndSepPunct{\mcitedefaultmidpunct}
{\mcitedefaultendpunct}{\mcitedefaultseppunct}\relax
\EndOfBibitem
\bibitem[Mardirossian and Head-Gordon(2016)Mardirossian, and
  Head-Gordon]{mardirossian_head_gordon_wb97m_v_functional_jcp_2016}
Mardirossian,~N.; Head-Gordon,~M. {$\omega$B97M-V: A Combinatorially Optimized,
  Range-Separated Hybrid, Meta-GGA Density Functional with VV10 Nonlocal
  Correlation}. \emph{J. Chem. Phys.} \textbf{2016}, \emph{144}, 214110\relax
\mciteBstWouldAddEndPuncttrue
\mciteSetBstMidEndSepPunct{\mcitedefaultmidpunct}
{\mcitedefaultendpunct}{\mcitedefaultseppunct}\relax
\EndOfBibitem
\bibitem[Hait and Head-Gordon(2018)Hait, and
  Head-Gordon]{hait_head_gordon_dipole_mom_jctc_2018}
Hait,~D.; Head-Gordon,~M. {How Accurate Is Density Functional Theory at
  Predicting Dipole Moments? An Assessment Using a New Database of 200
  Benchmark Values}. \emph{J. Chem. Theory Comput.} \textbf{2018}, \emph{14},
  1969\relax
\mciteBstWouldAddEndPuncttrue
\mciteSetBstMidEndSepPunct{\mcitedefaultmidpunct}
{\mcitedefaultendpunct}{\mcitedefaultseppunct}\relax
\EndOfBibitem
\bibitem[Sun \latin{et~al.}(2018)Sun, Berkelbach, Blunt, Booth, Guo, Li, Liu,
  McClain, Sayfutyarova, Sharma, Wouters, and Chan]{pyscf_wires_2018}
Sun,~Q.; Berkelbach,~T.~C.; Blunt,~N.~S.; Booth,~G.~H.; Guo,~S.; Li,~Z.;
  Liu,~J.; McClain,~J.~D.; Sayfutyarova,~E.~R.; Sharma,~S.; Wouters,~S.;
  Chan,~G. K.-L. {{\texttt{PySCF}}: The Python-Based Simulations of Chemistry
  Framework}. \emph{{W}IREs {C}omput. {M}ol. {S}ci.} \textbf{2018}, \emph{8},
  e1340\relax
\mciteBstWouldAddEndPuncttrue
\mciteSetBstMidEndSepPunct{\mcitedefaultmidpunct}
{\mcitedefaultendpunct}{\mcitedefaultseppunct}\relax
\EndOfBibitem
\bibitem[Sun \latin{et~al.}(2020)Sun, Zhang, Banerjee, Bao, Barbry, Blunt,
  Bogdanov, Booth, Chen, Cui, Eriksen, Gao, Guo, Hermann, Hermes, Koh, Koval,
  Lehtola, Li, Liu, Mardirossian, McClain, Motta, Mussard, Pham, Pulkin,
  Purwanto, Robinson, Ronca, Sayfutyarova, Scheurer, Schurkus, Smith, Sun, Sun,
  Upadhyay, Wagner, Wang, White, Whitfield, Williamson, Wouters, Yang, Yu, Zhu,
  Berkelbach, Sharma, Sokolov, and Chan]{pyscf_jcp_2020}
Sun,~Q.; Zhang,~X.; Banerjee,~S.; Bao,~P.; Barbry,~M.; Blunt,~N.~S.;
  Bogdanov,~N.~A.; Booth,~G.~H.; Chen,~J.; Cui,~Z.-H.; Eriksen,~J.~J.; Gao,~Y.;
  Guo,~S.; Hermann,~J.; Hermes,~M.~R.; Koh,~K.; Koval,~P.; Lehtola,~S.; Li,~Z.;
  Liu,~J.; Mardirossian,~N.; McClain,~J.~D.; Motta,~M.; Mussard,~B.;
  Pham,~H.~Q.; Pulkin,~A.; Purwanto,~W.; Robinson,~P.~J.; Ronca,~E.;
  Sayfutyarova,~E.~R.; Scheurer,~M.; Schurkus,~H.~F.; Smith,~J. E.~T.; Sun,~C.;
  Sun,~S.-N.; Upadhyay,~S.; Wagner,~L.~K.; Wang,~X.; White,~A.;
  Whitfield,~J.~D.; Williamson,~M.~J.; Wouters,~S.; Yang,~J.; Yu,~J.~M.;
  Zhu,~T.; Berkelbach,~T.~C.; Sharma,~S.; Sokolov,~A.~Y.; Chan,~G. K.-L.
  {Recent Developments in the {\texttt{PySCF}} Program Package}. \emph{J. Chem.
  Phys.} \textbf{2020}, \emph{153}, 024109\relax
\mciteBstWouldAddEndPuncttrue
\mciteSetBstMidEndSepPunct{\mcitedefaultmidpunct}
{\mcitedefaultendpunct}{\mcitedefaultseppunct}\relax
\EndOfBibitem
\bibitem[Lehtola \latin{et~al.}(2018)Lehtola, Steigemann, Oliveira, and
  Marques]{libxc_software_x_2018}
Lehtola,~S.; Steigemann,~C.; Oliveira,~M. J.~T.; Marques,~M. A.~L. {Recent
  Developments in {\texttt{LIBXC}} --- A Comprehensive Library of Functionals
  for Density Functional Theory}. \emph{Software X} \textbf{2018}, \emph{7},
  1\relax
\mciteBstWouldAddEndPuncttrue
\mciteSetBstMidEndSepPunct{\mcitedefaultmidpunct}
{\mcitedefaultendpunct}{\mcitedefaultseppunct}\relax
\EndOfBibitem
\bibitem[dec()]{decodense}
{\texttt{DECODENSE}}: A Decomposed Mean-Field Theory Code,
  See~{\url{https://github.com/januseriksen/decodense}}\relax
\mciteBstWouldAddEndPuncttrue
\mciteSetBstMidEndSepPunct{\mcitedefaultmidpunct}
{\mcitedefaultendpunct}{\mcitedefaultseppunct}\relax
\EndOfBibitem
\bibitem[Vydrov and Van~Voorhis(2010)Vydrov, and
  Van~Voorhis]{vydrov_voorhis_vv10_functional_jcp_2010}
Vydrov,~O.~A.; Van~Voorhis,~T. {Nonlocal van der Waals Density Functional: The
  Simpler the Better}. \emph{J. Chem. Phys.} \textbf{2010}, \emph{133},
  244103\relax
\mciteBstWouldAddEndPuncttrue
\mciteSetBstMidEndSepPunct{\mcitedefaultmidpunct}
{\mcitedefaultendpunct}{\mcitedefaultseppunct}\relax
\EndOfBibitem
\bibitem[Gill \latin{et~al.}(1993)Gill, Johnson, and
  Pople]{gill_johnson_pople_sg1_cpl_1993}
Gill,~P. M.~W.; Johnson,~B.~G.; Pople,~J.~A. {A Standard Grid for Density
  Functional Calculations}. \emph{Chem. Phys. Lett.} \textbf{1993}, \emph{209},
  506\relax
\mciteBstWouldAddEndPuncttrue
\mciteSetBstMidEndSepPunct{\mcitedefaultmidpunct}
{\mcitedefaultendpunct}{\mcitedefaultseppunct}\relax
\EndOfBibitem
\bibitem[Dasgupta and Herbert(2017)Dasgupta, and
  Herbert]{dasgupta_herbert_sg2_sg3_jcc_2017}
Dasgupta,~S.; Herbert,~J.~M. {Standard Grids for High-Precision Integration of
  Modern Density Functionals: SG-2 and SG-3}. \emph{J. Comput. Chem.}
  \textbf{2017}, \emph{38}, 869\relax
\mciteBstWouldAddEndPuncttrue
\mciteSetBstMidEndSepPunct{\mcitedefaultmidpunct}
{\mcitedefaultendpunct}{\mcitedefaultseppunct}\relax
\EndOfBibitem
\bibitem[Dunlap(2000)]{dunlap_densfit_pccp_2000}
Dunlap,~B.~I. {Robust and Variational Fitting}. \emph{Phys. Chem. Chem. Phys.}
  \textbf{2000}, \emph{2}, 2113\relax
\mciteBstWouldAddEndPuncttrue
\mciteSetBstMidEndSepPunct{\mcitedefaultmidpunct}
{\mcitedefaultendpunct}{\mcitedefaultseppunct}\relax
\EndOfBibitem
\bibitem[Not()]{Note-3}
The TIP3P/flex and AIMD/revPBE-D3 samplings both use TIP3P charge
  distributions, namely, $q(\text{O}) = -0.834$ and $q(\text{H}) = 0.417$,
  while the TIP4P/2005 sampling makes use of its own three-point model, i.e.,
  $q(\text{M}) = -1.1128$ and $q(\text{H}) = 0.5564$, with the negative charge
  placed on a dummy atom (M) at a distance of $0.1546$ {\AA} away from the
  oxygen along the $\angle$(H--O--H) bisector.\relax
\mciteBstWouldAddEndPunctfalse
\mciteSetBstMidEndSepPunct{\mcitedefaultmidpunct}
{}{\mcitedefaultseppunct}\relax
\EndOfBibitem
\bibitem[Bak{\'o} \latin{et~al.}(2019)Bak{\'o}, Daru, Pothoczki, Pusztai, and
  Hermansson]{bako_hermansson_water_dipole_jml_2019}
Bak{\'o},~I.; Daru,~J.; Pothoczki,~S.; Pusztai,~L.; Hermansson,~K. {Effects of
  H-bond Asymmetry on the Electronic Properties of Liquid Water --- An AIMD
  Analysis}. \emph{J. Mol. Liq.} \textbf{2019}, \emph{293}, 111579\relax
\mciteBstWouldAddEndPuncttrue
\mciteSetBstMidEndSepPunct{\mcitedefaultmidpunct}
{\mcitedefaultendpunct}{\mcitedefaultseppunct}\relax
\EndOfBibitem
\bibitem[Zhu \latin{et~al.}(2018)Zhu, de~Silva, and
  Van~Voorhis]{zhu_silva_voorhis_sah_decomp_jctc_2018}
Zhu,~T.; de~Silva,~P.; Van~Voorhis,~T. {Self-Attractive Hartree Decomposition:
  Partitioning Electron Density into Smooth Localized Fragments}. \emph{J.
  Chem. Theory Comput.} \textbf{2018}, \emph{14}, 92\relax
\mciteBstWouldAddEndPuncttrue
\mciteSetBstMidEndSepPunct{\mcitedefaultmidpunct}
{\mcitedefaultendpunct}{\mcitedefaultseppunct}\relax
\EndOfBibitem
\bibitem[Bak{\'o} and Mayer(2016)Bak{\'o}, and
  Mayer]{bako_mayer_water_dipole_jpca_2016}
Bak{\'o},~I.; Mayer,~I. {On Dipole Moments and Hydrogen Bond Identification in
  Water Clusters}. \emph{J. Phys. Chem. A} \textbf{2016}, \emph{120},
  4408\relax
\mciteBstWouldAddEndPuncttrue
\mciteSetBstMidEndSepPunct{\mcitedefaultmidpunct}
{\mcitedefaultendpunct}{\mcitedefaultseppunct}\relax
\EndOfBibitem
\bibitem[Kongsted \latin{et~al.}(2002)Kongsted, Osted, Mikkelsen, and
  Christiansen]{kongsted_water_dipole_cpl_2002}
Kongsted,~J.; Osted,~A.; Mikkelsen,~K.~V.; Christiansen,~O. {Dipole and
  Quadrupole Moments of Liquid Water Calculated within the Coupled
  Cluster/Molecular Mechanics Method}. \emph{Chem. Phys. Lett.} \textbf{2002},
  \emph{364}, 379\relax
\mciteBstWouldAddEndPuncttrue
\mciteSetBstMidEndSepPunct{\mcitedefaultmidpunct}
{\mcitedefaultendpunct}{\mcitedefaultseppunct}\relax
\EndOfBibitem
\bibitem[Kemp and Gordon(2008)Kemp, and Gordon]{gordon_water_dipole_jpca_2008}
Kemp,~D.~D.; Gordon,~M.~S. {An Interpretation of the Enhancement of the Water
  Dipole Moment Due to the Presence of Other Water Molecules}. \emph{J. Phys.
  Chem. A} \textbf{2008}, \emph{112}, 4885\relax
\mciteBstWouldAddEndPuncttrue
\mciteSetBstMidEndSepPunct{\mcitedefaultmidpunct}
{\mcitedefaultendpunct}{\mcitedefaultseppunct}\relax
\EndOfBibitem
\bibitem[Bak{\'o} and Mayer(2016)Bak{\'o}, and
  Mayer]{bako_mayer_water_dipole_jpcb_2016}
Bak{\'o},~I.; Mayer,~I. {Hierarchy of the Collective Effects in Water
  Clusters}. \emph{J. Phys. Chem. A} \textbf{2016}, \emph{120}, 631\relax
\mciteBstWouldAddEndPuncttrue
\mciteSetBstMidEndSepPunct{\mcitedefaultmidpunct}
{\mcitedefaultendpunct}{\mcitedefaultseppunct}\relax
\EndOfBibitem
\bibitem[Gregory \latin{et~al.}(1997)Gregory, Clary, Liu, Brown, and
  Saykally]{gregory_saykally_water_dipole_science_1997}
Gregory,~J.~K.; Clary,~D.~C.; Liu,~K.; Brown,~M.~G.; Saykally,~R.~J. {The Water
  Dipole Moment in Water Clusters}. \emph{Science} \textbf{1997}, \emph{275},
  814\relax
\mciteBstWouldAddEndPuncttrue
\mciteSetBstMidEndSepPunct{\mcitedefaultmidpunct}
{\mcitedefaultendpunct}{\mcitedefaultseppunct}\relax
\EndOfBibitem
\bibitem[Coulson and Eisenbergf(1966)Coulson, and
  Eisenbergf]{coulson_eisenbergf_ice_dipole_proc_soc_london_1966}
Coulson,~C.~A.; Eisenbergf,~D. {Interactions of H$_2$O Molecules in Ice I. The
  Dipole Moment of an H$_2$O Molecule in Ice}. \emph{Proc. R. Soc. London. Ser.
  A} \textbf{1966}, \emph{291}, 445\relax
\mciteBstWouldAddEndPuncttrue
\mciteSetBstMidEndSepPunct{\mcitedefaultmidpunct}
{\mcitedefaultendpunct}{\mcitedefaultseppunct}\relax
\EndOfBibitem
\bibitem[Batista \latin{et~al.}(1998)Batista, Xantheas, and
  J{\'o}nsson]{xantheas_jonsson_ice_dipole_jcp_1998}
Batista,~E.~R.; Xantheas,~S.~S.; J{\'o}nsson,~H. {Molecular Multipole Moments
  of Water Molecules in Ice Ih}. \emph{J. Chem. Phys.} \textbf{1998},
  \emph{109}, 4546\relax
\mciteBstWouldAddEndPuncttrue
\mciteSetBstMidEndSepPunct{\mcitedefaultmidpunct}
{\mcitedefaultendpunct}{\mcitedefaultseppunct}\relax
\EndOfBibitem
\bibitem[Liu \latin{et~al.}(2017)Liu, He, and
  Zhang]{liu_he_zhang_water_h_bonds_pccp_2017}
Liu,~J.; He,~X.; Zhang,~J. Z.~H. {Structure of Liquid Water --- A Dynamical
  Mixture of Tetrahedral and `Ring-and-Chain` Like Structures}. \emph{Phys.
  Chem. Chem. Phys.} \textbf{2017}, \emph{19}, 11931\relax
\mciteBstWouldAddEndPuncttrue
\mciteSetBstMidEndSepPunct{\mcitedefaultmidpunct}
{\mcitedefaultendpunct}{\mcitedefaultseppunct}\relax
\EndOfBibitem
\bibitem[Ruiz~Pestana \latin{et~al.}(2017)Ruiz~Pestana, Mardirossian,
  Head-Gordon, and Head-Gordon]{head_gordon_water_aimd_chem_sci_2017}
Ruiz~Pestana,~L.; Mardirossian,~N.; Head-Gordon,~M.; Head-Gordon,~T.
  {{\textit{Ab Initio}} Molecular Dynamics Simulations of Liquid Water Using
  High Quality Meta-GGA Functionals}. \emph{Chem. Sci.} \textbf{2017},
  \emph{8}, 3554\relax
\mciteBstWouldAddEndPuncttrue
\mciteSetBstMidEndSepPunct{\mcitedefaultmidpunct}
{\mcitedefaultendpunct}{\mcitedefaultseppunct}\relax
\EndOfBibitem
\bibitem[Riera \latin{et~al.}(2019)Riera, Lambros, G{\"o}tz, and
  Paesani]{paesani_water_h_bonds_chem_sci_2019}
Riera,~M.; Lambros,~T.~T.,~E.~Nguyen; G{\"o}tz,~A.; Paesani,~F. {Low-Order
  Many-Body Interactions Determine the Local Structure of Liquid Water}.
  \emph{Chem. Sci.} \textbf{2019}, \emph{10}, 8211\relax
\mciteBstWouldAddEndPuncttrue
\mciteSetBstMidEndSepPunct{\mcitedefaultmidpunct}
{\mcitedefaultendpunct}{\mcitedefaultseppunct}\relax
\EndOfBibitem
\bibitem[Torii(2013)]{torii_water_dipole_jpca_2013}
Torii,~H. {Extended Nature of the Molecular Dipole of Hydrogen-Bonded Water}.
  \emph{J. Phys. Chem. A} \textbf{2013}, \emph{117}, 2044\relax
\mciteBstWouldAddEndPuncttrue
\mciteSetBstMidEndSepPunct{\mcitedefaultmidpunct}
{\mcitedefaultendpunct}{\mcitedefaultseppunct}\relax
\EndOfBibitem
\bibitem[Badyal \latin{et~al.}(2000)Badyal, Saboungi, Price, Shastri, Haeffner,
  and Soper]{badyal_soper_water_exp_dipole_jcp_2000}
Badyal,~Y.~S.; Saboungi,~M.-L.; Price,~D.~L.; Shastri,~S.~D.; Haeffner,~D.~R.;
  Soper,~A.~K. {Electron Distribution in Water}. \emph{J. Chem. Phys.}
  \textbf{2000}, \emph{112}, 9206\relax
\mciteBstWouldAddEndPuncttrue
\mciteSetBstMidEndSepPunct{\mcitedefaultmidpunct}
{\mcitedefaultendpunct}{\mcitedefaultseppunct}\relax
\EndOfBibitem
\bibitem[Clough \latin{et~al.}(1973)Clough, Beers, Klein, and
  Rothman]{clough_rothman_water_exp_dipole_jcp_1973}
Clough,~S.~A.; Beers,~Y.; Klein,~G.~P.; Rothman,~L.~S. {Dipole Moment of Water
  from Stark Measurements of H$_2$O, HDO, and D$_2$O}. \emph{J. Chem. Phys.}
  \textbf{1973}, \emph{59}, 2254\relax
\mciteBstWouldAddEndPuncttrue
\mciteSetBstMidEndSepPunct{\mcitedefaultmidpunct}
{\mcitedefaultendpunct}{\mcitedefaultseppunct}\relax
\EndOfBibitem
\bibitem[Lide(2004)Lide]{crc_85}
Lide,~D.~R. {CRC Handbook of Chemistry and Physics}. 
85th ed.; CRC Press, 2004\relax
\mciteBstWouldAddEndPuncttrue
\mciteSetBstMidEndSepPunct{\mcitedefaultmidpunct}
{\mcitedefaultendpunct}{\mcitedefaultseppunct}\relax
\EndOfBibitem
\bibitem[Unke and Meuwly(2019)Unke, and
  Meuwly]{unke_meuwly_ml_physnet_jctc_2019}
Unke,~O.~T.; Meuwly,~M. {{\texttt{PhysNet}}: A Neural Network for Predicting
  Energies, Forces, Dipole Moments, and Partial Charges}. \emph{J. Chem. Theory
  Comput.} \textbf{2019}, \emph{15}, 3678\relax
\mciteBstWouldAddEndPuncttrue
\mciteSetBstMidEndSepPunct{\mcitedefaultmidpunct}
{\mcitedefaultendpunct}{\mcitedefaultseppunct}\relax
\EndOfBibitem
\bibitem[Veit \latin{et~al.}(2020)Veit, Wilkins, Yang, DiStasio~Jr., and
  Ceriotti]{ceriotti_ml_dipole_jcp_2020}
Veit,~M.; Wilkins,~D.~M.; Yang,~Y.; DiStasio~Jr.,~R.~A.; Ceriotti,~M.
  {Predicting Molecular Dipole Moments by Combining Atomic Partial Charges and
  Atomic Dipoles}. \emph{J. Chem. Phys.} \textbf{2020}, \emph{153},
  024113\relax
\mciteBstWouldAddEndPuncttrue
\mciteSetBstMidEndSepPunct{\mcitedefaultmidpunct}
{\mcitedefaultendpunct}{\mcitedefaultseppunct}\relax
\EndOfBibitem
\bibitem[L.-P. \latin{et~al.}(2013)L.-P., Head-Gordon, Ponder, Ren, Chodera,
  Eastman, Martinez, and Pande]{pande_water_ff_jpcb_2013}
Wang,~L.-P.; Head-Gordon,~T.; Ponder,~J.~W.; Ren,~P.; Chodera,~J.~D.;
  Eastman,~P.~K.; Martinez,~T.~J.; Pande,~V.~S. {Systematic Improvement of a
  Classical Molecular Model of Water}. \emph{J. Phys. Chem. B} \textbf{2013},
  \emph{117}, 9956\relax
\mciteBstWouldAddEndPuncttrue
\mciteSetBstMidEndSepPunct{\mcitedefaultmidpunct}
{\mcitedefaultendpunct}{\mcitedefaultseppunct}\relax
\EndOfBibitem
\bibitem[Lambros and Paesani(2020)Lambros, and
  Paesani]{paesani_water_ff_jcp_2020}
Lambros,~E.; Paesani,~F. {How Good are Polarizable and Flexible Models for
  Water: Insights from a Many-Body Perspective}. \emph{J. Chem. Phys.}
  \textbf{2020}, \emph{153}, 060901\relax
\mciteBstWouldAddEndPuncttrue
\mciteSetBstMidEndSepPunct{\mcitedefaultmidpunct}
{\mcitedefaultendpunct}{\mcitedefaultseppunct}\relax
\EndOfBibitem
\end{mcitethebibliography}
\end{document}